\documentstyle[12pt,epsf,epsfig,here]{article}

\newcommand{\AmS}{{\protect\the\textfont2
  A\kern-.1667em\lower.5ex\hbox{M}\kern-.125emS}}
\def\beq{\begin{equation}}
\def\eeq{\end{equation}}
\def\bea{\begin{eqnarray}}
\def\beaa{\begin{eqnarray*}}
\def\eea{\end{eqnarray}}
\def\eeaa{\end{eqnarray*}}
\def\bq{\begin{quote}}
\def\eq{\end{quote}}
\def\gappeq{\mathrel{\rlap {\raise.5ex\hbox{$>$}}
{\lower.5ex\hbox{$\sim$}}}}

\begin{document}
\pagestyle{empty}
\begin{flushright}
CERN-TH/99-294\\
{\tt hep-th/9909156}
\end{flushright}
\vspace*{1cm}
\begin{center}
{\bf Confining Properties of Abelian-Projected Theories and \\
Field Strength Correlators}\\
\vspace*{1cm}
{\bf D. Antonov}\\
Institute of Theoretical and Experimental Physics,
\\
        RU - 117 218 Moscow, Russia\\
\vspace*{0.3cm}
        and\\
\vspace*{0.3cm}
       {\bf D. Ebert}\\
Theoretical Physics Division, CERN, \\
        CH - 1211 Geneva 23 \\ and \\
Institut f\"ur Physik, Humboldt-Universit\"at,\\D - 10115 Berlin, Germany\\
\vspace*{1cm}

ABSTRACT
\end{center}

We review the string representations of Abelian-projected SU(2)-
and SU(3)-gauge theories and their application to the evaluation
of bilocal field strength correlators. The large distance
asymptotic behaviours of the latter ones are shown to be in agreement
with the Stochastic Vacuum Model of QCD and existing lattice data.

\vspace*{0.3cm}
\begin{center}
{\it Invited talk given at the }\\
{\it Euroconference ``QCD 99"}\\
{\it 7-13 July 1999}\\
{\it Montpellier, France }
\end{center}
\vspace*{0.2cm}

\begin{flushleft}
CERN-TH/99-294\\
September 1999
\end{flushleft}
\vfill
\eject
\pagestyle{plain}
\section{INTRODUCTION}
The explanation and description of the confinement phenomenon
in gauge theories is known to be one of the most
fundamental challenges of the modern QFT (see
{\it e.g.}~\cite{polbook}).
Till now, an analytical description of this phenomenon is best of all
elaborated on in the theories containing monopole ensembles.
Those include, in particular, 3D compact QED~\cite{polbook}, where
monopoles form a dilute gas, and QCD-inspired
Abelian-projected $SU(N)$ gauge theories~\cite{tH}.
In the latter case, the
dominance of Abelian degrees of freedom is usually
realized by assuming that non-Abelian (charged) gauge bosons are heavy and
do not propagate at a long-distance scale. In particular by partially fixing
the gauge, one removes there as many non-Abelian
degrees of freedom as possible,
leaving  the maximal Abelian subgroup  $[U(1)]^{N-1}$ unbroken.
The relevant IR degrees of freedom of the resulting
Abelian-projected theories are then described by the field strengths of
the $(N-1)$ Abelian (neutral) gauge bosons supplemented by the same amount
of magnetic Dirac strings  generated by singular gauge
transformations.  It is further very convenient to reformulate
these theories by means of a duality
transformation,
which leads to a theory of dual Abelian gauge fields interacting with
$(N-1)$ patterns of monopole currents.
Finally, performing the summation over the
grand canonical
ensembles of such currents by making use of the
Bardakci-Samuel formula~\cite{Bard}, one obtains an equivalent
effective Ginzburg-Landau type Lagrangian for disorder magnetic Higgs
fields. Those belong to
the maximal Abelian subgroup $[U(1)]^{N-1}$~\cite{Suz} and describe the
condensates of monopole Cooper pairs.
The effective theory described by this Lagrangian
allows for the formation of nonvanishing
{\it v.e.v.}'s
of magnetic Higgs fields mentioned above
and thus realizes the beautiful dual Meissner scenario of confinement
proposed many years ago by
`t Hooft and Mandelstam~\cite{tHMand}.

Since confinement is usually associated with the formation of strings
(tubes of electric flux between external quarks)~\cite{wils},
it seems natural to seek for the description of this phenomenon in
terms of elementary string excitations. Obviously, the dynamical scheme
of a dual superconductor mentioned above
provides us with a very convenient theoretical framework for studying
this problem by a derivation of string representations of the
related effective Abelian-projected theories.
In fact, due to the multivaluedness of the phase of the Higgs field,
there should arise singularities in the dual gauge fields,
which are natural to be identified
with   electric Abrikosov-Nielsen-Olesen (ANO) type
strings~\cite{ano}. Moreover, by virtue of the Higgs mechanism,
monopole condensation leads
to the nonvanishing mass of the dual gauge
bosons, which sets the scale of the resulting string tension.

In the present talk, we shall briefly review recent progress achieved
in the construction of string representations of the
above mentioned Abelian-projected $SU(N)$ theories.
Possible consequences of the obtained results
for the realistic QCD, based on the so-called Stochastic Vacuum
Model (SVM)~\cite{svm}, will also be discussed.
In our interpretation of this subject in the two subsequent sections,
where the $SU(2)$- and $SU(3)$-cases will be successively considered,
we shall follow
Refs.~\cite{epj} and~\cite{plb} (see Ref.~\cite{dis} for
an extended review).

\section{SU(2)--THEORY}

As it is commonly argued, the
Abelian-projected $SU(2)$-gluodynamics
is just the
Dual Abelian Higgs Model (DAHM), whose action
in the London limit ({\it i.e.} the limit of infinitely large
coupling constant $\lambda$ of the magnetic Higgs field) has the form
\bea
S_{``SU(2)"}=
\int d^4x\left[
\frac14 F_{\mu\nu}^2 +
\frac{\eta^2}{2}
\left(\partial_\mu\theta - 2g B_\mu\right)^2\right].
\label{1}
\eea
Here, $F_{\mu\nu} = \partial_\mu B_\nu - \partial_\nu B_\mu$ is the field
strength tensor  of the dual vector potential $B_\mu$, $g$ is the
magnetic coupling constant, and
$\eta$ is the
{\it v.e.v.} of the magnetic Higgs field.
In the London limit under study, the radial part of the latter one
has been integrated out, whereas its phase has the form
$\theta = \theta^{\rm sing} + \theta^{\rm reg}$,
where the multivalued part
$\theta^{\rm sing}$
describes a given electric
string configuration, while $\theta^{\rm reg}$ stands for a single-valued
fluctuation around such a configuration. Owing to the fact that
the singularities of the phase of the magnetic Higgs field occur
at the world-sheets of closed electric
ANO type strings~\cite{ano},
there exists a  correspondence between $\theta^{\rm sing}$
and string world-sheets, given by the equation
\bea
\varepsilon_{\mu\nu\lambda\rho}\partial_\lambda\partial_\rho
\theta^{\rm sing}(x)=2\pi\Sigma_{\mu\nu}(x)\equiv 
2\pi\int
\limits_{\Sigma}^{}d\sigma_{\mu\nu}(x(\xi))\delta(x-x(\xi)).
\label{2}
\eea
Here, $x(\xi)\equiv x_\mu(\xi)$ is a vector parametrizing the
world-sheet $\Sigma$ with $\xi=\left(\xi^1,\xi^2\right)$ standing
for the 2D coordinate. This correspondence eventually enables one
to reformulate the integration over $\theta^{\rm sing}$'s as an
integration over $x_\mu(\xi)$'s. 
The resulting partition function  has the form
${\cal Z}_{``SU(2)"} = \int {\cal D} x_\mu(\xi)$  $\exp\left(
-S_{\rm str.}^{``SU(2)"}\right)$,
where the string effective action reads \cite{epj}
\bea
S_{\rm str.}^{``SU(2)"}=\frac{g\eta^3}{2}
\int d^4x\int d^4y 
\Sigma_{\mu\nu}(x)
\frac{K_1(m|x-y|)}{|x-y|}\Sigma_{\mu\nu}(y).
\label{3}
\eea
Here, $m=2g\eta$ is the mass of the dual gauge boson generated by
the Higgs mechanism, and $K_1$ stands for the modified Bessel
function.
The reader is referred to the above cited papers for details of the
so-called path-integral duality transformation, which leads to
Eq.~(\ref{3}).

Performing the expansion of the action~(\ref{3}) in powers of the
derivatives {\it w.r.t.}
$\xi^a$'s, it has
been shown that the first two terms of this expansion are the standard
Nambu-Goto one and the so-called rigidity term, {\it i.e.}
\bea
S_{\rm str.}^{``SU(2)"}\simeq
\sigma\int d^2\xi\sqrt{\hat g}+
\frac{1}{\alpha_0}
\int d^2\xi\sqrt{\hat g}\hat g^{ab}\left(\partial_a t_{\mu\nu}\right)
\left(\partial_b t_{\mu\nu}\right).
\label{4}
\eea
Here,
$\partial_a=\partial/\partial\xi^a$,
$\hat g=\det \left|\left|\hat g^{ab}\right|\right|$
with $\hat g^{ab}=(\partial^a x_\mu(\xi))
(\partial^b x_\mu(\xi))$ being the induced metric tensor of the
world-sheet, and $t_{\mu\nu}=\frac{\varepsilon^{ab}}{\sqrt{\hat g}}
\left(\partial_a x_\mu(\xi)\right)\left(\partial_b x_\nu(\xi)\right)$
standing for the  extrinsic curvature tensor.
The coupling constant of the
Nambu-Goto term (also called string tension)
with the logarithmic accuracy reads $\sigma \simeq \pi\eta^2\ln
\frac{\sqrt{\lambda}}{g}$, while
the inverse coupling constant of the rigidity term (considered first in Ref.
\cite{PolKlei}) has the form
$\frac{1}{\alpha_0}=-\frac{\pi}{32g^2}$.
In particular, if
external quarks are introduced into the system,
the Nambu-Goto action yields a linearly rising quark-antiquark
potential
$V_{\rm conf}(R)=\sigma R$.
Notice also the negative sign of the coupling $\alpha_0$,
which reflects the stability of strings.

Another important subject investigated in Ref.~\cite{epj}
is the evaluation of the irreducible bilocal field strength correlator
(cumulant) $\left<\left<\tilde F_{\lambda\nu}(x)
\tilde F_{\mu\rho}(0)\right>\right>$, where
$\tilde F_{\mu\nu}\equiv\frac12
\varepsilon_{\mu\nu\lambda\rho}F_{\lambda\rho}$, by a derivation
of its string representation. Parametrizing the $x_\mu$-independent
Lorentz structure of the cumulant according to the
SVM~\cite{svm} as
$\left(\delta_{\lambda\mu}\delta_{\nu\rho}-
\delta_{\lambda\rho}\delta_{\nu\mu}\right)
{\cal D}\left(x^2\right)$,
we find the following IR asymptotic behaviour of the
function ${\cal D}$ at the distances $|x|\gg \frac{1}{m}$

\begin{equation}
\label{d}
{\cal D}\to\frac{m^4}{4\sqrt{2}\pi^{\frac32}}\frac{
{\rm e}^{-m|x|}}{(m|x|)^{\frac32}}.
\end{equation}
This behaviour is very similar to the one observed in the lattice
simulations of QCD in Ref.~\cite{lat}. In particular, one can see that
the r\^ole of the so-called correlation length of the vacuum, $T_g$,
at which the cumulant in SVM decreases, is played in
DAHM by the inverse mass of the dual gauge boson, $m^{-1}$.
The evaluation of the second coefficient function
parametrizing the cumulant, which stands at the omitted $x_\mu$-dependent
Lorentz structure, also matches the existing lattice data~\cite{lat} with
a good accuracy.

\section{SU(3)--THEORY}

The effective Abelian-projected theory of the
$SU(3)$-gluodynamics~\cite{Suz} is also of the DAHM type,
albeit with the $[U(1)]^2$ gauge invariance  {\it w.r.t.} the
maximal Abelian subgroup of the $SU(3)$-group. In the London limit,
the action under study reads

\bea
S_{``SU(3)"}=
\int d^4x\left[\frac14\vec
F_{\mu\nu}^2+\frac{\eta^2}{2}
\sum\limits_{a=1}^{3}\left(\partial_\mu\theta_a-2g\vec\varepsilon_a
\vec B_\mu\right)^2\right],
\label{5}
\eea
where $\vec F_{\mu\nu}=\partial_\mu\vec B_\nu-\partial_\nu\vec B_\mu$
is the field strength tensor of magnetic fields $\vec B_\mu\equiv
\left(B_\mu^3, B_\mu^8\right)$, which are dual to the usual gluonic fields
$A_\mu^3$ and $A_\mu^8$. Next, in Eq.~(\ref{5}) the so-called root vectors
$\vec\varepsilon_1=(1,0)$,
$\vec\varepsilon_2=\left(-\frac12,-\frac{\sqrt{3}}{2}\right)$,
$\vec\varepsilon_3=\left(-\frac12,\frac{\sqrt{3}}{2}\right)$ have been
introduced. These vectors naturally emerge during the Cartan decomposition as
the structural constants in the commutation relations between the
diagonal and (properly redefined) non-diagonal $SU(3)$-generators.
Besides that, the action~(\ref{5}) is assumed to be supplied by the
following constraint imposed on the phases $\theta_a$ of magnetic
Higgs fields,
$\sum\limits_{a=1}^{3}\theta_a=0$, which
is just the reflection of the fact that the original
$SU(3)$ group is special. Next,
the relation~(\ref{2}) remains
the same, with the substitution $\theta^{\rm sing}\to\theta_a^{\rm sing}$,
$\Sigma_{\mu\nu}\to\Sigma_{\mu\nu}^a$, $\Sigma\to\Sigma^a$, and
$x(\xi)\to x^{(a)}(\xi)$. Consequently, there exist three different
types of electric strings, among which, however, only two are independent
of each other owing to the above-mentioned constraint.
Performing the path-integral duality transformation of the
theory~(\ref{5}) and integrating out
one of the world-sheets (for concreteness, $x_\mu^{(3)}$), we arrive
at the desired string representation for the partition function,
which reads ${\cal Z}_{``SU(3)"}=\int {\cal D}x_\mu^{(1)}(\xi)
{\cal D} x_\mu^{(2)}(\xi)\exp\left(-S_{\rm str.}^{``SU(3)"}\right)$.
Here, the string effective action has the form~\cite{plb}

$$
S_{\rm str.}^{``SU(3)"}=g\eta^3\sqrt{\frac32}\int d^4x\int d^4y\times
$$

\begin{equation}
\times\left[\Sigma_{\mu\nu}^1(x)\Sigma_{\mu\nu}^1(y)+
\Sigma_{\mu\nu}^1(x)\Sigma_{\mu\nu}^2(y)
+\Sigma_{\mu\nu}^2(x)\Sigma_{\mu\nu}^2(y)\right]
\frac{K_1\left(m_B|x-y|\right)}{|x-y|},
\label{6}
\end{equation}
where $m_B=\sqrt{6}g\eta$ is the mass of the fields $\vec B$. One can see
that, according to Eq.~(\ref{6}), the most crucial difference of the
string effective theory corresponding to the Abelian-projected
$SU(3)$-gluodynamics  {\it w.r.t.} the $SU(2)$-case is the presence
of two independent kinds of strings, which not only self-interact,
but also interact with each other by the exchanges of the massive dual
gauge bosons.

As far as the cumulants of the field strength tensors
$\tilde F_{\mu\nu}^{3,8}$ are concerned, only for those of them,
which are the diagonal ones, {\it i.e.}
$\left<\left<\tilde F_{\lambda\nu}^3(x)
\tilde F_{\mu\rho}^3(0)\right>\right>$ and
$\left<\left<\tilde F_{\lambda\nu}^8(x)
\tilde F_{\mu\rho}^8(0)\right>\right>$, the vacuum does exhibit a
nontrivial correlation length $T_g=\frac{1}{m_B}$. In particular,
the IR asymptotics~(\ref{d})
of the function ${\cal D}$
remains valid for these two cumulants with the
replacement $m\to m_B$.

\section{CONCLUSIONS}
The obtained
results demonstrate the relevance
of the method of Abelian projection and the path-integral duality
transformation to the description of the
confining properties of the $SU(2)$- and $SU(3)$-gluodynamics.
The approach considered above also provides us with a new
field-theoretical status of SVM and sheds some light on the
problem of finding an adequate string representation of QCD.
Finally, it is worth mentioning that the field strength correlators
in DAHM have been also investigated beyond the London limit
in Refs.~\cite{beyond} and~\cite{dis}.

\vspace{3mm}

\noindent
{\bf Discussions}

\vspace{5mm}

\noindent
{\bf N. Brambilla} (University of Vienna)

\textit{Why do you use the London limit?
I think that this limit is appropriate for
large transverse distances
from the string, but not for large distances between an
external quark and an antiquark.}
\vspace*{0.5cm}

\noindent
{\bf D. Ebert}

\textit{The London limit has been used
here as a simplifying assumption leading to
infinitely thin strings. If one includes
an external quark-antiquark pair,
this scheme leads to a confinement
potential plus a Yukawa interaction (arising
as a boundary term).
Indeed, it would be interesting to go beyond the London
limit in the sense of taking into account
the vanishing of the Higgs field
inside the finite core of strings. This is expected to yield, besides the usual
confinement potential, a Coulomb potential instead of the Yukawa one.}


\vspace*{1cm}



\begin{thebibliography}{20}
\bibitem{polbook}
A.M. Polyakov, {\it Gauge Fields and Strings}
(Harwood Academic, Chur, 1987).


\bibitem{tH}
G. 't Hooft, Nucl. Phys. {\bf B 190} (1981) 455;
A.S. Kronfeld, G. Schierholz, and U.-J. Wiese,
Nucl. Phys. {\bf B 293} (1987) 461.


\bibitem{Bard}
K. Bardakci and S. Samuel, Phys. Rev. {\bf D 18} (1978) 2849.


\bibitem{Suz}
S. Maedan and T. Suzuki, Prog. Theor. Phys.
{\bf 81} (1989) 229.


\bibitem{tHMand}
G. 't Hooft, in: {\it High Energy Physics}, ed. A. Zichichi
(Editrice Compositori, Bologna, 1976);
S. Mandelstam, Phys. Rep. {\bf C 23} (1976) 245.


\bibitem{wils}
K.G. Wilson, Phys. Rev. {\bf D 10} (1974) 2445.

\bibitem{ano}
A.A. Abrikosov, Sov. Phys.- JETP {\bf 5} (1957) 1174;
H.B. Nielsen and P. Olesen, Nucl. Phys. {\bf B 61} (1973) 45.

\bibitem{svm}
H.G. Dosch, Phys. Lett. {\bf B 190} (1987) 177;
Yu.A. Simonov, Nucl. Phys. {\bf B 307} (1988) 512;
H.G. Dosch, Prog. Part. Nucl. Phys. {\bf 33} (1994) 121;
Yu.A. Simonov, Phys. Usp. {\bf 39} (1996) 313.

\bibitem{epj}
D. Antonov and D. Ebert, Eur. Phys. J. {\bf C 8} (1999) 343.

\bibitem{plb}
D. Antonov and D. Ebert, Phys. Lett. {\bf B 444} (1998) 208.

\bibitem{dis}
D. Antonov,
Ph.D. thesis at the Humboldt University
of Berlin (1999), available under {\tt http://dochost.rz.hu-berlin.de}.

\bibitem{PolKlei}
A. M. Polyakov, Nucl. Phys. {\bf B 268} (1986) 406;
H. Kleinert, Phys. Lett. {\bf B 174} (1986) 335.

\bibitem{lat}
A. Di Giacomo and H. Panagopoulos, Phys. Lett.
{\bf B 285} (1992) 133; A. Di Giacomo, E. Meggiolaro, and H. Panagopoulos,
Nucl. Phys. {\bf B 483} (1997) 371; A. Di Giacomo, M. D'Elia,
H. Panagopoulos, and E. Meggiolaro, preprint {\tt hep-lat/9808056}
(1998).


\bibitem{beyond}
M. Baker, N. Brambilla, H.G. Dosch, and A. Vairo, Phys. Rev. {\bf D 58}
(1998) 034010.
\end{thebibliography}
\end{document}